\documentclass{aa}  

\usepackage{graphicx}

\usepackage[breaklinks=true, citecolor=green]{hyperref}
\usepackage{txfonts}
\usepackage{xcolor}

\begin{document}

   \title{Exploring the physical origins of halo assembly bias from early times}

   \titlerunning{The physical origins of halo assembly bias}

    \authorrunning{Montero-Dorta et al.}

   \author{Antonio D. Montero-Dorta\inst{1}\thanks{antonio.montero@usm.cl}, Sergio Contreras\inst{2}, M. Celeste Artale\inst{3}, Facundo Rodriguez\inst{4,5} \& Ginevra Favole\inst{6, 7}
           }

   \institute{
             Departamento de F\'isica, Universidad T\'ecnica Federico Santa Mar\'ia, Avenida Vicu\~na Mackenna 3939, San Joaqu\'in, Santiago, Chile   
             \and Donostia International Physics Center, Manuel Lardizabal Ibilbidea, 4, 20018 Donostia, Gipuzkoa, Spain \and Universidad Andres Bello, Facultad de Ciencias Exactas, Departamento de Ciencias F\'isicas, Instituto de Astrof\'isica, Av. Fern\'andez Concha 700, Santiago, Chile \and CONICET. Instituto de Astronomía Teórica y Experimental (IATE). Laprida 854, Córdoba X5000BGR, Argentina \and Universidad Nacional de Córdoba (UNC). Observatorio Astronómico de Córdoba (OAC). Laprida 854, Córdoba X5000BGR, Argentina
             \and Instituto de Astrofísica de Canarias, s/n, E-38205, La Laguna, Tenerife, Spain
             \and 
             Departamento de Astrof\'{\i}sica, Universidad de La Laguna, E-38206, La Laguna, Tenerife, Spain
             }

   \date{Received --; accepted --}

 
  \abstract
   {The large-scale linear halo bias encodes the relation between the clustering of dark-matter (DM) halos and that of the underlying matter density field. Although the primary dependence of bias on halo mass is well understood in the context of structure formation, the physical origins of the multiple additional relations at fixed halo mass, commonly known as secondary halo bias, have not been fully elucidated. Of particular relevance is the secondary dependence on halo assembly history, known as halo assembly bias.  
   }
    {Our goal is to determine whether the properties of the initial regions from which $z=0$ halos originate produce any secondary bias at $z=0$. By analyzing these initial dependencies in connection with halo assembly bias, we intend to provide insight on the physical origins of the effect.
    }
  {We select halos at $z=0$ in the IllustrisTNG DM-only simulation and trace back the positions and velocities of their DM particles to $z=12$. The resulting initial regions are characterized according to several shape-related and kinematic properties. The secondary bias signal produced by these properties at $z=0$ is measured using an object-by-object bias estimator, which offers significant analytical advantages as compared to the traditional approach.}
   {We show that, when split by the properties of their initial DM clouds, $z=0$ halos display significant secondary bias, clearly exceeding the amplitude of the well-known halo assembly bias signal produced by concentration and age. The maximum bias segregation is measured for cloud velocity dispersion and radial velocity, followed by cloud concentration, sphericity, ellipticity and triaxiality. We further show that both velocity dispersion and radial velocity are also the properties of the initial clouds that most strongly correlate with halo age and concentration at fixed halo mass. Our results highlight the importance of linear effects in shaping halo assembly bias.}
   {}


   \maketitle
%

\section{Introduction}

The large-scale linear bias of dark matter (DM) halos encodes fundamental information on the process of halo formation from the underlying matter density field. Halo bias depends primarily on the peak height of density fluctuations, $\nu$, as it can be analytically derived from $\Lambda$-cold dark matter ($\Lambda$-CDM) structure formation formalisms (see, e.g.,\citealt{Kaiser1984,Bardeen1986,Mo1996, ShethTormen1999, Sheth2001, Tinker2010}). This fundamental dependence naturally manifests itself on halo virial mass, which dictates that more massive halos are more tightly clustered (have higher bias) than less massive halos. In the last decades, this rather simple picture has become more complex, as a number of additional secondary dependencies of halo bias (i.e., at fixed halo mass) have been unveiled using cosmological simulations (see, e.g., \citealt{Sheth2004,gao2005,Wechsler2006,Gao2007,Dalal2008, Angulo2008,Li2008,faltenbacher2010, Lazeyras2017,2018Salcedo,Mao2018, Han2018,SatoPolito2019, Johnson2019, Ramakrishnan2019,Contreras2019, MonteroDorta2020B, Tucci2021,  Contreras2021_cosmo, MonteroDorta2021, MonteroRodriguez2024, Balaguera2024}). Among these dependencies, which are called {\it{secondary halo bias}} by some authors, the dependence on the assembly history of halos, or {\it{halo assembly bias}}, is the effect that has drawn more attention: at fixed halo mass, halos that assemble a significant fraction of their mass earlier on are more tightly clustered than those that take more time to acquire the same mass. Throughout this work, halo assembly bias is specifically defined as the dependence on two internal assembly properties: virial concentration and half-mass formation scale factor\footnote{Note that multiple definitions of halo age have been analyzed in this context, as discussed by \citealt{Li2008}.} (the scale factor at which the halo have acquired half its final $z=0$ mass).

The physical origins of the multiple secondary halo bias dependencies are still not fully determined, although several key results have been reported. For halo assembly bias, following \cite{Dalal2008}, it is common to divide the study into two different mass regimes, motivated by the conspicuous reversal of the secondary-bias signal measured for concentration at $M_{\rm vir}\simeq 10^{13}$ $h^{-1}$M$_\odot$ (e.g., \citealt{Gao2007, SatoPolito2019}). At the high-mass end, \cite{Dalal2008} proposed that 
the fact that lower-concentration halos have higher bias than higher-concentration halos was tightly connected with the statistics of density peaks in the initial field. At fixed $\nu$, peaks of lower curvature are more clustered and tend to accrete their mass earlier on as compared to high-curvature peaks, which suggests a connection with the concentration dependence of halo bias.    

The above argument is clearly not valid at the low-mass end, as pointed out by \cite{Dalal2008} as well (it is also unclear how the peak-curvature mechanism would be adapted to other proxies of assembly history, such as half-mass formation time, for which no signal of secondary bias is measured at the high-mass end). For these low-mass halos, there are indications that assembly bias may be influenced by non-linear effects\footnote{The term ``non-linear" here refers to physical mechanisms that involve other halos, i.e., collapsed structures.} that may depend on the location of halos within the cosmic web. These effects can hinder mass accretion early on for a subpopulation of halos \citep{Dalal2008}. The suppression of halo growth may be produced by tidal effects induced individually by a neighboring massive halo (e.g., \citealt{Hahn2009}) or, more generally, by the whole geometry of tides \citep{Borzyszkowski2017, Musso2018}. Pinpointing the exact mechanisms is nevertheless complicated, given the large intrinsic stochasticity in bias, and the complex map of correlations between environmental and internal halo properties. A clearer conceptual picture is to view low-mass assembly bias as a consequence of correlations between bias and the anisotropy of the local tidal field, which in turn correlates with the internal halo properties at fixed halo mass (\citealt{Paranjape2018, Ramakrishnan2019}). Note that this picture can be adapted to include other related environmental properties as mediators of the effect (see \citealt{Balaguera2024, MonteroRodriguez2024}). 

The results reported in the literature provide valuable clues to the physical origins of halo assembly bias. However, a comprehensive theory is yet to be established. In particular, it is still unclear how the different physical mechanisms that have been proposed, both in the linear and non-linear regimes, contribute to the signals across different halo mass ranges. In this paper, we aim to provide new insights by explicitly connecting halos at $z=0$ with the initial density field. In practice, we use the IllustrisTNG\footnote{\url{http://www.tng-project.org}} DM-only simulation to investigate the properties of the collapsing regions at $z=12$ where different types of halos form. These properties are analyzed in connection with the halo assembly bias signal measured at $z=0$, in order to shed light on the processes that shape this effect. To facilitate the analysis of bias and its correlations with multiple properties, an object-by-object halo bias estimator is employed (\citealt{Paranjape2018, Contreras2021a,Balaguera2024}).  

The paper is organized as follows. Section \ref{sec:data} provides a brief description of the simulation data used in this work. Our methodology for measuring the properties of the $z=12$ initial clouds and the $z=0$ halo-by-halo bias is described in Section \ref{sec:method}. Our main results regarding assembly bias are presented in Section \ref{sec:results}. Finally, Section \ref{sec:conclusions} is devoted to discussing the implications of our results and providing a brief summary of the paper. The IllustrisTNG300 simulation adopts the standard $\Lambda$CDM cosmology \citep{Planck2016}, with parameters $\Omega_{\rm m} = 0.3089$,  $\Omega_{\rm b} = 0.0486$, $\Omega_\Lambda = 0.6911$, $H_0 = 100\,h\, {\rm km\, s^{-1}Mpc^{-1}}$ with $h=0.6774$, $\sigma_8 = 0.8159$, and $n_s = 0.9667$.

\section{Data}
\label{sec:data}

Our analysis is based on publicly available data from the IllustrisTNG suite of cosmological simulations (hereafter, TNG, for simplicity, \citealt{Pillepich2018b,Pillepich2018,Nelson2018_ColorBim,Nelson2019,Marinacci2018,Naiman2018,Springel2018}). The TNG simulations, particularly the magnetohydrodynamical boxes, have been employed in a large number of works over the years, providing insight on a variety of galaxy formation processes (see https://www.tng-project.org/results/ for a complete list of publications). Since we are only interested in DM halos, we analyze the TNG300-1 Dark box (hereafter, TNG300-DMO\footnote{\url{https://www.tng-project.org/data/docs/specifications/}}, for DM-only), which follows the dynamical evolution of 2500$^3$ DM particles of mass $5.0 \times 10^7$ $h^{-1} {\rm M_{\odot}}$ within a simulated cosmological volume of $205\,\,h^{-1}$Mpc (with periodic boundary conditions). Choosing the largest box within the TNG suite provides an obvious advantage when it comes to measuring large-scale halo clustering, as we will discuss in Section \ref{sec:method}. DM halos in TNG are identified using a friends-of-friends (FOF) algorithm with a linking length of 0.2 times the mean inter-particle separation \citep{Davis1985}. The {\sc subfind} algorithm \citep{Springel2001,Dolag2009} is in turn used to identify subhalos.  

Several halo properties at $z=0$ are employed in this work. We use the virial halo mass, $M_{\rm vir}$ [$h^{-1} {\rm M_{\odot}}$], defined as the total mass enclosed within a sphere of radius $R_{\rm vir}$, where the density equals 200 times the critical density of the Universe. From $M_{\rm vir}$ and the merger trees provided in the database, we compute the standard proxy for halo age, i.e., the half-mass scale factor $a_{\rm form}$, which represents the time at which half the $z=0$ halo mass $M_{\rm vir}$ has formed. We also measure the virial concentration, $c_{\rm vir}$, defined in the standard way as the ratio between the virial radius and the scale radius, i.e.,  $c_{\rm vir} = R_{\rm vir}/R_{\rm s}$ (where $R_{\rm s}$ is obtained by fitting the DM density profiles of individual halos to a NFW profile, \citealt{nfw1997}). 

As a proxy for the initial field, we use the $z=12$ snapshot. This is the earliest (in terms of cosmic time) complete (full-particle) snapshot in TNG300-DMO. In the following section, we describe the set of properties measured from the collapsing regions that are identified in this snapshot as the progenitors of $z=0$ halos.  

\section{Methodology}
\label{sec:method}

\begin{figure}
	\includegraphics[width=1\columnwidth]{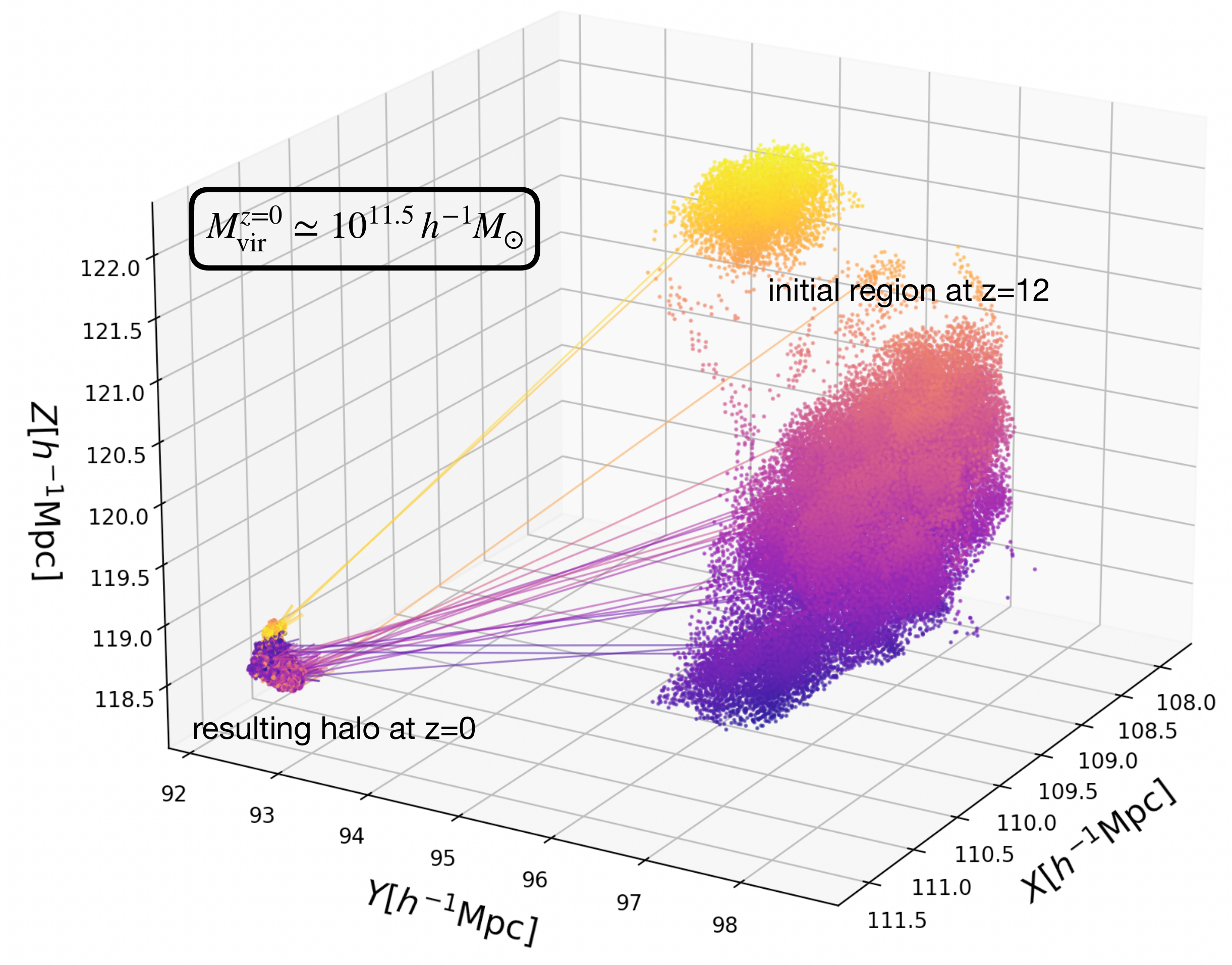}
    \caption{A collapsing region at $z=12$ and the corresponding halo that forms from that patch of the density field, viewed at $z=0$ . In this figure, 10 $\%$ of the particles are plotted for simplicity. The color code, from purple to yellow, simply indicates the height in the z-axis at $z=12$, from low to high. Arrows indicating the initial and final location of a few individual DM particles are shown. Note that this particular region is composed of two disjointed sub-regions that are in the process of collapsing and merging.}
    \label{fig:halos}
\end{figure}

\begin{figure*}
    \centering
    \includegraphics[width=2\columnwidth]{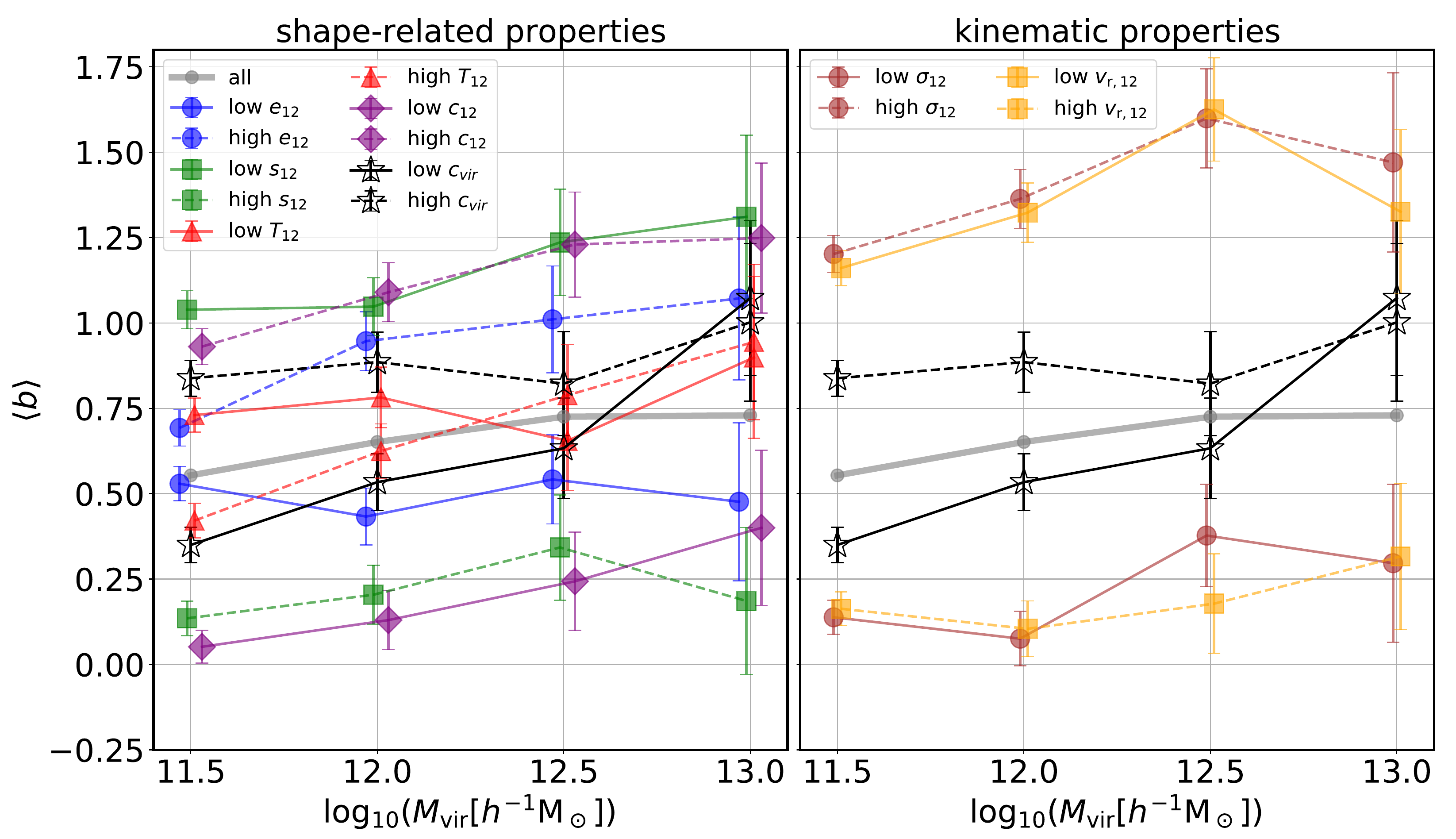}
    \caption{The secondary halo bias (measured at $z=0$) produced by a set of properties related to the shape (left) and kinematics (right) of the collapsing regions at $z=12$ from which $z=0$ halos in specific mass bins originate. These results are compared with the assembly bias signal, represented here by the concentration, $c_{\rm vir}$, secondary dependence. The mean biases in percentile subsets encompassing 20$\%$ (low) and 80$\%$ (high) of the halo population in each mass bin, along the mean bias of the entire population in that mass bin, are shown as a function of halo mass. Errors are estimated using a bootstrap technique based on 1000 resampling subsets. The lines connecting data have been added only for the sake of a better visualization of the differences between subsets.}
    \label{fig:assembly_bias_1}
\end{figure*}

The main goal of this work is to analyze the secondary bias signal (i.e., the differences in bias at fixed halo mass) produced by the properties of the $z=12$ regions from which $z=0$ halos form. Our procedure can be divided into the following simple steps:   

\begin{enumerate}
    \item  We select all halos at $z=0$ within 4 discontinuous narrow halo mass bins of width 0.2 dex and centered at $\log_{10 }M_{\rm vir}[h^{-1} \rm M_\odot]=$ [11.5, 12, 12.5, 13]. This selection ensures that differences in bias are measured at approximately fixed halo mass. 
    \item We trace the DM particles composing each of the selected halos all the way back to $z=12$ (snapshot number 2 in TNG300-DMO).
    \item With the positions and velocities of the selected particles, several internal properties of the resulting $z=12$ clouds are measured. As we describe below, we focus on velocity dispersion, radial velocity, sphericity, ellipticity, triaxiality, and mass concentration. 
    \item We measure the secondary bias signal produced by the internal properties of the initial regions. 
    The $z=0$ halo population in each mass bin is split using these properties, thus explicitly establishing a relation between the high-$z$ regions and the $z=0$ halos. The large-scale linear bias is computed at $z=0$.

\end{enumerate}

Fig. \ref{fig:halos} shows an example of a collapsing region at $z=12$ and the resulting virialized object (halo) of mass $M_{\rm vir} = 10^{11.5} \, h^{-1}{\rm M_\odot}$ at $z=0$. In the remainder of this section, we describe in more detail the calculation of the $z=12$ properties, and the individual halo-by-halo bias at $z=0$.

\subsection{Characterizing collapsing regions at $z=12$}

The following properties of the $z=12$ regions are measured based on the positions and velocities of traced-back particles: 

\begin{itemize}

    \item velocity dispersion ($\sigma_{12}$ [km/s]): the square root of the variance of the velocity of particles with respect to the mean velocity of the region.  
    
    \item radial velocity ($v_{r, 12}$ [km/s]): the velocity dispersion of the region projected onto the radial direction, which is defined with respect to the mean velocity of the region. Negative values of this property correspond to inward bulk motions.    
    
    \item sphericity ($s_{12}$): computed by diagonalizing the inertia tensor, so that the principal axes of the ideal ellipsoid can be obtained from the eigenvalues as $a=\sqrt{\lambda_1}$, $b=\sqrt{\lambda_2}$, $c=\sqrt{\lambda_3}$, with $a>b>c$. The sphericity is defined as $s_{12}=c/a$, so larger values mean more spherical objects.
    
    \item ellipticity ($e_{12}$): computed from the principal axes of the ellipsoid as $e_{12} = (1-s_{12}^2)/(1+s_{12}^2+q_{12}^2)$, where $s_{12}$ is the sphericity and $q_{12}=b/a$. Larger values indicate more elliptical objects. 
    
    \item triaxiality ($T_{12}$): computed from the principal axes of the ellipsoid as $T_{12} = (1-q_{12}^2)/(1-s_{12}^2)$. Values of $T\lesssim 1/3$ are usually associated with oblate shapes, whereas objects with $T\gtrsim 2/3$ are considered prolate. The range $1/3 \lesssim T \lesssim 2/3$ corresponds to triaxial objects not clearly fitting into either of these categories. These objects have a more complex shape, with all three axes having different lengths.
    
    \item mass concentration ($c_{12}$): measured as the radius that contains 90$\%$ of the particles divided by the radius that contains 50$\%$ of the particles, following the philosophy of the virial concentration defined for halos. As in that case, larger values of $c_{12}$ correspond to more concentrated clouds.

\end{itemize}

The above properties are sometimes measured within our framework for regions of very complex structure. In some cases, these regions are composed of multiple sub-regions. This is illustrated in Fig. \ref{fig:halos}, where the initial region is composed of 2 disjointed sub-regions that are in the process of collapsing and merging. The above properties, therefore, provide an average characterization of the shape and kinematics of the whole patch, which is naturally  dominated by the most massive components. In follow-up work, we will refine our measurement by characterizing the multiple initial components individually. 

\begin{figure*}
    \centering
    \includegraphics[width=1.9\columnwidth]{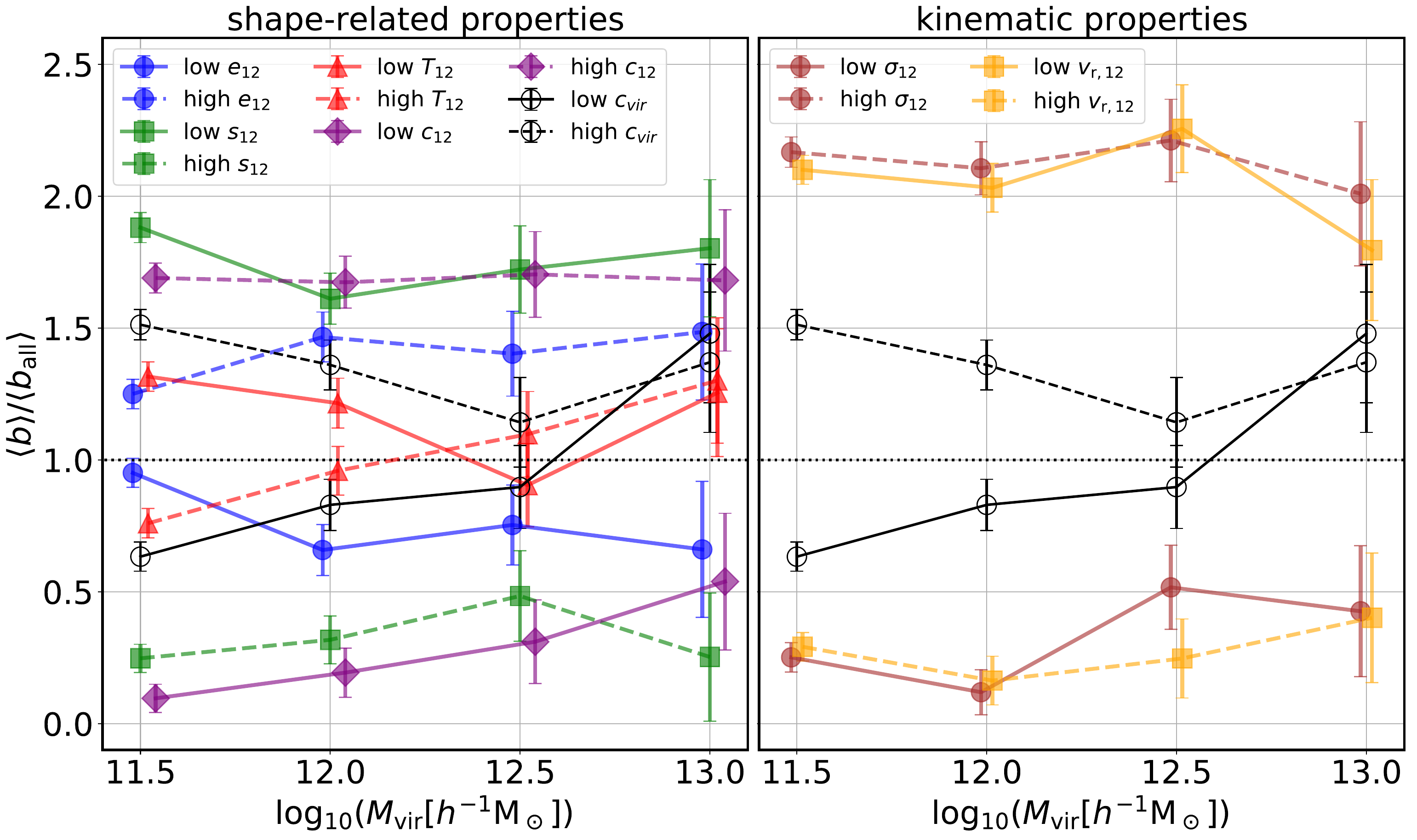}
    \caption{Similar to Fig. \ref{fig:assembly_bias_1} but showing the mean relative bias as a function of halo mass for the same set of properties, which are related to the shape (left) and kinematics (right) of the initial clouds. The mean relative bias is simply defined as the ratio between the mean bias of the low-/high-percentile subset and the mean bias of the entire mass bin. Errors are estimated using a bootstrap procedure based on 1000 resampling subsets. The lines connecting data points are simply added to facilitate the readability of the plot.}
    \label{fig:assembly_bias_2}
\end{figure*}

\begin{figure*}
    \centering
    \includegraphics[width=2\columnwidth]{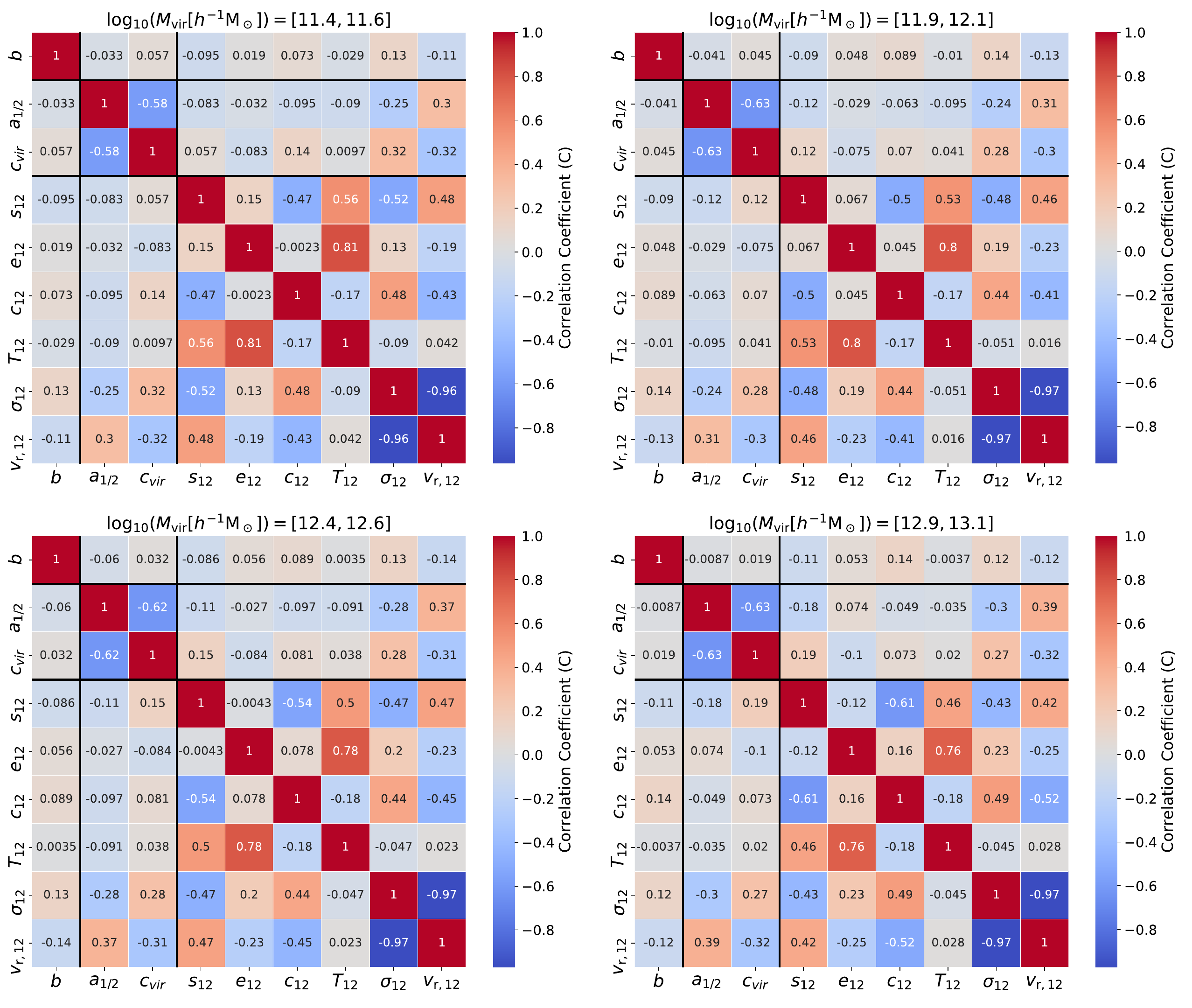}
    \caption{The correlation matrix for all properties analyzed in this work in each of the 4 halo mass bins considered. Three different groups of properties are highlighted with thicker lines: clustering (bias), internal $z=0$ halo properties and the properties of the initial $z=12$ clouds.}
    \label{fig:correlation_matrix}
\end{figure*}

\begin{figure*}
    \centering
    \includegraphics[width=1.9\columnwidth]{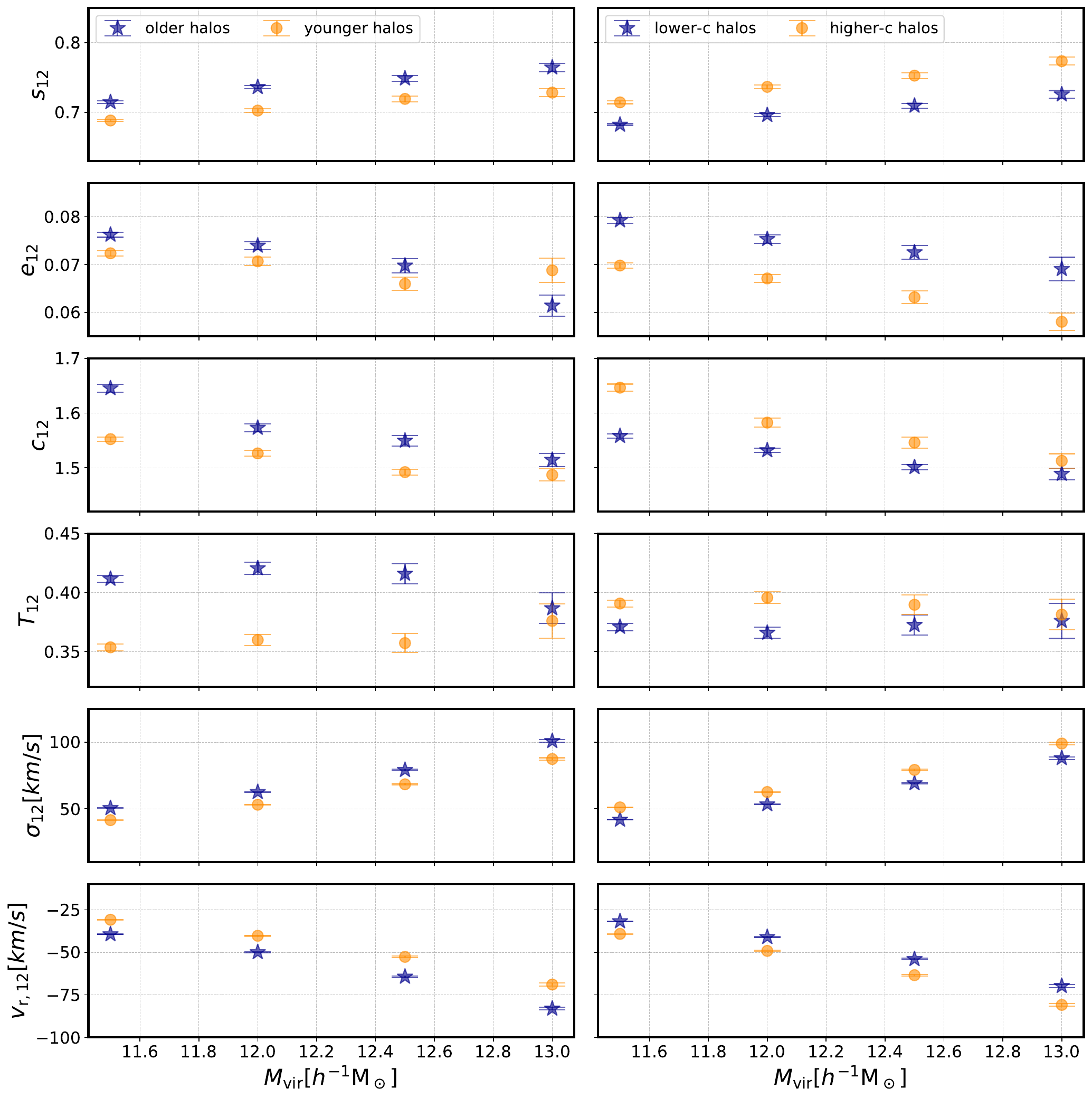}
    \caption{Left: The average properties of the $z=12$ collapsing regions from which older (low $a_{1/2}$) and younger (high $a_{1/2}$) halos originate. Right: The same for virial concentration (low and high $c_{\rm vir}$). From top to bottom, the mean values of sphericity, ellipticity, concentration, triaxiality, velocity dispersion and radial velocity of the corresponding clouds are presented for each subset separately. Errors are estimated using a bootstrap procedure based on 1000 resampling subsets.}
    \label{fig:correlation_all}
\end{figure*}

\subsection{Measuring secondary bias}

The standard way of measuring linear halo bias is based on a ratio of either correlation functions or power spectra, e.g., $b = \xi_{hm}/\xi_{mm}$ (where $\xi_{mm}$ is the auto-correlation of the DM density field and $\xi_{hm}$ is the cross-correlation between halos and the DM itself). These procedures have the disadvantage that the bias is measured for subsets of halos, which can complicate some analyses that aim at disentangling correlations between internal halo properties, environment and bias. Instead, we use the object-by-object estimate of halo bias developed by \cite{Paranjape2018}. In this prescription, the effective large-scale bias is viewed as a sample mean in which each element contributes with an object-by-object bias, $b_i$, of the form:

\begin{equation}
    b_i = b^{(i)}_{hm}=\frac{\sum_{j,k_{j}<k_{max}}N^{j}_{k}\langle \exp[-i\bf{k}\cdot \bf{r}_{i}]\delta_{\mathrm{DM}}^{*}(\bf{k}) \rangle_{k_{j}}}{\sum_{j,k_{j}<k_{max}} N^{j}_{k}P_{\rm DM}},
   \label{eq:bias}
\end{equation}

\noindent where $\delta_{\mathrm{DM}}(\bf{k})$ is the Fourier transform of the DM density field, $P_{\rm DM}$ is the matter power spectrum, and $N^{j}_{k}$ is the number of Fourier modes in the j-th spherical shell. The sum is carried over the range of wavenumbers in which the ratio between the halo and the DM power spectra is constant. We have taken such limit as $0.04 \leq k[h \, \rm{Mpc}^{-1}] \leq 0.08$. For computational feasibility, 1/256 of the total number of particles of the DM field are employed. For similar applications of the individual bias, see \cite{Ramakrishnan2019, Contreras2021a, Balaguera2024}.

With the object-by-object bias in hand, measuring secondary bias is trivial. In each selected mass bin, the halo set can be split by a particular property (internal at $z=0$ or from the original $z=12$ region) using percentile subsets encompassing 20 and 80$\%$ of the sample. Subsequently, the mean individual bias of each subset is computed and an error on the mean is assigned based on a bootstrap method. This is the traditional approach, but in order to harness the power of the halo-by-halo bias, the measurement can be also inserted into a correlation analysis as another individual halo property.

\section{Results}
\label{sec:results}

Fig. \ref{fig:assembly_bias_1} displays the secondary halo bias produced at $z=0$ by several properties of the collapsing regions (measured at $z=12$), as compared to the assembly bias signal. For simplicity, only the $c_{\rm vir}$ trend is plotted for assembly bias, since $a_{1/2}$ is known to produce a similar signal within the halo mass range considered. In both panels, the mean halo-by-halo bias is presented as a function of the $z=0$ halo mass, showing the characteristic enhancement of clustering towards the high-mass end (see, e.g., \citealt{Tinker2010} for a detailed discussion and fitting functions). Note that lines connecting data have been added only for the sake of a better visualization of the differences between subsets. As mentioned in the previous section, different percentile subsets are created by ranking and splitting the halo population within each mass bin based on the values of each property (20$\%$, 80$\%$ and the entire sample). Errors throughout this work have been estimated by means of a bootstrap analysis, based on 1000 resampling subsets. Secondary bias exists when a significant difference is measured between the mean bias of the low- and the high-percentile subsets. Note that, as long as the bias has a monotonous dependence on a given property, the mean bias of the entire population must lie between the extreme subsets. 

Fig. \ref{fig:assembly_bias_1} shows one of the key results of this work: when properties of the initial collapsing regions are assigned to the resulting halos at $z=0$, large secondary bias signals, often exceeding in amplitude the assembly bias trend, are obtained. In order to better quantify this result, in Fig. \ref{fig:assembly_bias_2} we show the relative bias produced by the same properties, i.e., the ratio between the mean bias of each percentile subset and the mean bias of the entire population in that mass bin. From the analysis of Figs. \ref{fig:assembly_bias_1} and \ref{fig:assembly_bias_2}, the following conclusions are drawn:

\begin{itemize}

    \item Below $\log_{10} M_{\rm vir}[h^{-1} \rm M_\odot] \simeq 13$, halos at $z=0$ that originate from collapsing regions at $z=12$ that have higher concentration, lower sphericity, higher ellipticity, higher velocity dispersion and more negative (i.e., higher, in  absolute values) radial velocity are more clustered on large scales than halos in the opposite subsets. Lower triaxiality (more oblate) halos seem to also have a higher bias towards the low-mass end. 

    \item Among the above trends, the largest secondary bias signals within the aforementioned mass range are found for velocity dispersion and radial velocity, with mean relative biases of $>2$ ($<0.25$) for the higher- (lower-) bias subsets. Also, concentration and sphericity display large signals, with factors of $\sim 1.75$ ($\sim0.25$) in relative bias. Ellipticity also yields a signal that exceeds that of virial concentration (factors 1.5-0.3). The secondary bias produced by triaxiality, on the other hand, is only significant, albeit smaller than the concentration trend, for low-mass halos (1.3-0.75).

    \item All properties, except for triaxiality, produce signals that, although strong, show very little dependence on mass (even at the high mass end). This is different from the well-known assembly bias trends, which vanish  
    towards the high-mass end (both for concentration and age, within the halo mass range considered). The triaxiality trend appears to be the one that more closely resembles the concentration trend, with vanishing signal towards the high-mass end.

\end{itemize}

The results presented in Figs. \ref{fig:assembly_bias_1} and \ref{fig:assembly_bias_2} are not obvious. They imply that, at fixed $z=0$ halo mass, (individual) $z=0$ halo bias correlates strongly with the properties of the initial regions where halos originate. These figures also highlight two specific properties that display the strongest dependencies: the radial velocity and the velocity dispersion. These properties are tightly anti-correlated, since the radial velocity is a projection of the velocity dispersion and is negative in regions that are collapsing.  

The fact that selecting halos by different properties produces segregation in bias is a consequence of often small, although non-negligible correlations (or anti-correlations) between bias and these properties at fixed halo mass. In order to investigate the extent of these correlations and to provide a complete picture, Fig. \ref{fig:correlation_matrix} displays the correlation matrices for bias, and the $z=0$ and $z=12$ properties in the 4 halo mass bins considered. There are several important conclusions that we can extract from this figure. First of all, it shows that, due to the large scatter in all relations, the amount of correlation needed for a significant secondary bias signal to exist is rather small. The largest correlation (in absolute values) with bias is measured, as expected, for the velocity dispersion and the radial velocity of the collapsing cloud, with values of the correlation coefficient of $C \simeq +/-0.14$, respectively. It is interesting that several properties display almost null correlation coefficients with respect to the individual bias, and still show significant secondary bias. 

It is also obvious from Fig. \ref{fig:correlation_matrix} that significant correlation exists within the $z=0$ and the $z=12$ subset of properties. Virial concentration and age are, as expected, strongly anti-correlated throughout the entire mass range ($C \simeq -0.6$). Among the high-$z$ properties, it is also not surprising that the pairs that display stronger correlation are $e_{12}-T_{12}$ and $v_{r,12}-\sigma_{12}$: $C \simeq 0.8$ and $C \simeq -0.97$, respectively. Importantly, the high-$z$ properties that are more tightly connected with halo concentration and age are also $v_{r,12}$ and $\sigma_{12}$ in all halo mass bins, which suggests, when analyzed in the context of bias, that these are the properties of the initial regions that more directly shape the assembly bias trend. 

To complement the above analysis, Fig. \ref{fig:correlation_all} displays the average characteristics of the $z=12$ regions from which higher/lower-concentration and older/younger $z=0$ halos originate. Note that this figure reflects the correlations displayed in Fig. \ref{fig:correlation_all}, but provides additional information on the actual values of each property. On average, higher-$c_{\rm vir}$ halos originate in regions of higher sphericity, and consequently lower ellipticity, as compared to low-$c_{\rm vir}$ halos. Those regions are also more concentrated and characterized by higher velocity dispersions and more negative radial velocities. Similar results are obtained for older halos (as expected from the $c_{\rm vir}$-$a_{1/2}$ anti-correlation), although some interesting discrepancies exist. For example, older halos appear to be more spherical, but at the same time slightly more elliptical than younger halos. This contradiction might reflect the structural complexity of the initial regions (recall that these regions can be composed of disjointed patches). Fig. \ref{fig:correlation_all} also shows that virial concentration is a better tracer of the original ellipticity, whereas age discriminate better between regions of different triaxiality. 

\section{Discussion and Conclusions}
\label{sec:conclusions}

In this work, we trace the positions and velocities of the DM particles that make up a selected subset of $z=0$ halos all the way back to $z=12$. In particular, all halos in 4 specific narrow mass bins, covering the mass range $10^{11.5} < M_{\rm vir}[h^{-1}\rm M_\odot] < 10^{13}$ at $z=0$, are considered in the analysis. The goal of the paper is to examine the properties of the initial regions that give rise to these objects (initial in the sense that structure formation is still in its linear regime), and to evaluate the degree of secondary bias that these properties display at $z=0$. This analysis is intended to shed light on the buildup of the assembly bias signal (the secondary bias dependence on halo assembly history, i.e., either on concentration or formation time) from the initial density field. 

We have shown that halo concentration and formation time retain information on the properties of the initial regions. In terms of statistical correlation, the velocity dispersion and radial velocity of the initial clouds have the highest correlation coefficients with respect to these internal halo properties ($C \pm 0.25-0.35$, in absolute values), followed by sphericity ($C \pm 0.10-0.2$). The rest of the properties display values of $C \lesssim 0.1$. In general, as we have checked looking at the mean values of the properties of the initial regions, older/more concentrated halos at fixed halo mass tend to originate, as compared to younger/less concentrated halos, in more spherical and concentrated patches, characterized also by higher velocity dispersions and radial velocities (in amplitude, meaning more negative values of the latter). Although these regions are fairly spherical, older/concentrated halos also appear to emerge from slightly more oblate regions than those where younger/less concentrated halos form.  

Importantly, using a convenient halo-by-halo estimate of bias at $z=0$, we have shown that several properties of the initial regions display large secondary bias signals, when they are used to split the $z=0$ halo population at fixed halo mass. This is a clear indication of the importance of linear effects in shaping assembly bias., i.e., those that can be directly connected with properties of the initial density field. The shapes of the trends themselves, however, might potentially reflect the impact of non-lineal effects that become important at later stages of halo evolution. We will come back to this point later in this section. 

Halos with the same mass that formed in regions of, in this order, higher (more negative) velocity dispersion (radial velocity), lower sphericity, higher concentration and higher ellipticity are significantly more clustered than the opposite subpopulations, in extents that exceed considerably the assembly bias trends. Only triaxiality seems to display a weaker secondary dependence than halo concentration and halo age. This hierarchy is reflected in the correlation matrix, where we can include the bias thanks to the individual nature of the estimate. The correlation between these initial properties and bias tends to be, as expected, stronger than for halo properties, e.g., $\sim$ 0.13 for $\sigma_{12}$, as compared to $\sim$ 0.05 for concentration. Note that it is well known that small correlations with bias can produce large secondary bias signals (see, e.g., \citealt{MonteroDorta2020B, Paranjape2018}). Importantly, our results highlight a clear link between the velocity dispersion and radial velocity of the initial regions and assembly bias, which can be easily understood from a physical standpoint. The radial velocity, $v_{\rm r}$, for example, is related to peculiar velocities, which respond to the gravitational potential of the region, thus reflecting the ``strength" of the collapse. This picture is consistent with our result that older/more concentrated halos have typically more negative values of $v_{\rm r}$, indicating that these objects tend to collapse faster. Note that, since the absolute value of the radial velocity is highly correlated with the velocity dispersion, it is also possible to understand this bias trend in terms of virial mass alone. The regions that give rise to early-formed halos will tend to reach a higher virial mass earlier on than late-formed halos, even though they both reach the same virial mass at $z=0$. It is reasonable, therefore, that these regions are more biased at early times. 

The large amplitudes of secondary bias measured in this paper for the initial velocity dispersion and the radial velocity have only been previously found for environmental properties. These are not internal properties, but they have been proposed as mediators of the assembly bias effect. Examples of this can be found in \cite{Paranjape2018}, where clustering is measured as a function of the anisotropy of the environment (the so-called $\alpha$ parameter) at fixed halo mass, or in \cite{MonteroRodriguez2024}, where a similar analysis is performed as a function of the distance to the main cosmic-web structures (characterized by the critical points of the density field). By exploring the correlation with properties of the initial collapsing clouds, we have been able to identify simple ``internal" quantities (in the sense that they can be computed from the same set of particles that make up a current halo) that maximize the secondary bias signal at $z=0$.     

As mentioned above, another interesting aspect to consider is the shape of the secondary bias trends measured. Whereas both assembly bias signals decrease with mass up to $M_{\rm halo}\simeq 10^{13} \, h^{-1}{\rm M_\odot}$, most of the trends related to the initial clouds show almost no dependence with mass, thus maintaining the same signal at the high-mass end (we will not considered the well-known reversal of the signal above this mass scale for concentration, since that is beyond the mass range analyzed here). The cause of the specific mass dependence of assembly bias is unclear. The drop in signal towards $M_{\rm vir}\simeq 10^{13} \, h^{-1} {\rm M_\odot}$ could be the imprint of non-linear effects that act at a later stage of halo evolution. The correlations between different properties of the initial regions might also provide some insight. As the correlations between different properties display a dependence on mass, the vanishing of the assembly bias signal could potentially be related to the cancellation of different dependencies. As an example, the correlation between concentration and sphericity increases significantly with halo mass. Sphericity is anti-correlated with bias, whereas velocity dispersion correlates with both concentration and bias. Trends like these could provide hints on the physical mechanisms that make the assembly bias trend vanish at the high-mass end. Unfortunately, the small volume of TNG300 prevents us from properly investigating this hypothesis. We will address this aspect in a follow-up work, where we analyze multiple snapshots in a bigger cosmological volume, which will allow us to reach halo masses around $M_{\rm vir}\simeq 10^{14} \, h^{-1} {\rm M_\odot}$ at high statistical significance. 

Our analysis can be improved by taking into account that, when DM halos are traced back, the resulting initial regions are sometimes composed of two or potentially more disjointed sub-regions, which are in the process of merging. Characterizing the properties of these sub-regions separately, along with their relation with the properties of virialized objects at $z=0$, will add another layer of complexity to our analysis. In this context, the study of the build-up of the assembly bias signal can be extended by evaluating the connection between the initial regions and halos at different redshifts. 

Finally, a potential application of the properties investigated in this work is the inclusion of assembly bias in galaxy population models. Models such as Halo Occupation Distributions (HODs, \citealt{Jing1998, Benson2000, Peacock2000, Berlind2003, Zheng2005, Zheng2007, Contreras2013, Guo2015, Contreras2017}) and SubHalo Abundance Matching (SHAM, \citealt{Vale2006, Conroy2006, Reddick2013, Contreras2015, Guo2016, ChavesMontero2016, Lehmann2017, Dragomir2018,  Artale2018, Hadzhiyska2021, Favole2022}) contain a limited or even zero level of assembly bias (we refer to galaxy assembly bias in this case\footnote{Here, we are implicitly adopting the definition of galaxy assembly bias based on so-called {\it{occupancy variations}}, i.e., the dependence of halo occupation on the assembly history of halos and related quantities.}). Some extensions of these models have been able to inject additional assembly bias to improve their accuracy. Initially, these models add assembly bias to their galaxies by correlating the occupation number or galaxy properties (such as stellar mass or luminosity) to a halo property that displays a certain level of halo assembly bias (e.g., \citealt{Hearin2015}). The amount of galaxy assembly bias this method could add was limited by the amount of halo assembly bias the halo property had, which was usually insufficient \citep{Xu2021b}. Note that here we are specifically referring to properties such as concentration or formation time, thus the use of the term assembly bias. Modern mocks add bias using environmental properties, a type of secondary bias which is not technically assembly bias but has a similar (actually stronger) effect \citep{Paranjape2018, Contreras2021a, Contreras2021b, Xu2021a}. Using the properties found in this work could help introduce assembly bias in a more natural way, which could in turn facilitate the use of these mocks on a variety of halo-galaxy connection studies.

\begin{acknowledgements}

This work is the result of the ideas and analyses developed during the Institute for Fundamental Physics of the Universe (IFPU) Team Research Program {\it{Unveiling the physical origins of assembly bias}}, which was held in Trieste in July 2024. We thank IFPU for their financial support and Emiliano Sefusatti and Pierluigi Monaco for their hospitality and valuable discussions during the week. ADMD and FR also acknowledge the Abdus Salam International Centre for Theoretical Physics (ICTP) for their hospitality and financial support through the Senior Associates Programme 2022-2027 and Junior Associates Programme 2023-2028, respectively. 
ADMD thanks Fondecyt for financial support through the Fondecyt Regular 2021 grant 1210612. FR thanks the support by Agencia Nacional de Promoci\'on Cient\'ifica y Tecno\'ologica, the Consejo Nacional de Investigaciones Cient\'{\i}ficas y T\'ecnicas (CONICET, Argentina) and the Secretar\'{\i}a de Ciencia y Tecnolog\'{\i}a de la Universidad Nacional de C\'ordoba (SeCyT-UNC, Argentina). 
MCA acknowledges support from FONDECYT Iniciaci\'on 2024 grant number 11240540 and ANID BASAL project FB210003. SC acknowledges the support of the `Juan de la Cierva Incorporac\'ion' fellowship (IJC2020-045705-I).
\end{acknowledgements}

%
%

\bibliographystyle{aa} 
\bibliography{references} 

\begin{thebibliography}{69}
\expandafter\ifx\csname natexlab\endcsname\relax\def\natexlab#1{#1}\fi

\bibitem[{{Angulo} {et~al.}(2008){Angulo}, {Baugh}, \& {Lacey}}]{Angulo2008}
{Angulo}, R.~E., {Baugh}, C.~M., \& {Lacey}, C.~G. 2008, \mnras, 387, 921

\bibitem[{{Artale} {et~al.}(2018){Artale}, {Zehavi}, {Contreras}, \& {Norberg}}]{Artale2018}
{Artale}, M.~C., {Zehavi}, I., {Contreras}, S., \& {Norberg}, P. 2018, \mnras, 480, 3978

\bibitem[{{Balaguera-Antolinez} \& {Montero-Dorta}(2024)}]{Balaguera2024}
{Balaguera-Antolinez}, A. \& {Montero-Dorta}, A.~D. 2024, arXiv e-prints, arXiv:2407.09282

\bibitem[{{Bardeen} {et~al.}(1986){Bardeen}, {Bond}, {Kaiser}, \& {Szalay}}]{Bardeen1986}
{Bardeen}, J.~M., {Bond}, J.~R., {Kaiser}, N., \& {Szalay}, A.~S. 1986, \apj, 304, 15

\bibitem[{{Benson} {et~al.}(2000){Benson}, {Cole}, {Frenk}, {Baugh}, \& {Lacey}}]{Benson2000}
{Benson}, A.~J., {Cole}, S., {Frenk}, C.~S., {Baugh}, C.~M., \& {Lacey}, C.~G. 2000, \mnras, 311, 793

\bibitem[{{Berlind} {et~al.}(2003){Berlind}, {Weinberg}, {Benson}, {Baugh}, {Cole}, {Dav{\'e}}, {Frenk}, {Jenkins}, {Katz}, \& {Lacey}}]{Berlind2003}
{Berlind}, A.~A., {Weinberg}, D.~H., {Benson}, A.~J., {et~al.} 2003, \apj, 593, 1

\bibitem[{Borzyszkowski {et~al.}(2017)Borzyszkowski, Porciani, Romano-Díaz, \& Garaldi}]{Borzyszkowski2017}
Borzyszkowski, M., Porciani, C., Romano-Díaz, E., \& Garaldi, E. 2017, MNRAS, 469, 594–611

\bibitem[{{Chaves-Montero} {et~al.}(2016){Chaves-Montero}, {Angulo}, {Schaye}, {Schaller}, {Crain}, {Furlong}, \& {Theuns}}]{ChavesMontero2016}
{Chaves-Montero}, J., {Angulo}, R.~E., {Schaye}, J., {et~al.} 2016, \mnras, 460, 3100

\bibitem[{{Conroy} {et~al.}(2006){Conroy}, {Wechsler}, \& {Kravtsov}}]{Conroy2006}
{Conroy}, C., {Wechsler}, R.~H., \& {Kravtsov}, A.~V. 2006, \apj, 647, 201

\bibitem[{{Contreras} {et~al.}(2021{\natexlab{a}}){Contreras}, {Angulo}, \& {Zennaro}}]{Contreras2021a}
{Contreras}, S., {Angulo}, R.~E., \& {Zennaro}, M. 2021{\natexlab{a}}, \mnras, 504, 5205

\bibitem[{{Contreras} {et~al.}(2021{\natexlab{b}}){Contreras}, {Angulo}, \& {Zennaro}}]{Contreras2021b}
{Contreras}, S., {Angulo}, R.~E., \& {Zennaro}, M. 2021{\natexlab{b}}, \mnras, 508, 175

\bibitem[{{Contreras} {et~al.}(2013){Contreras}, {Baugh}, {Norberg}, \& {Padilla}}]{Contreras2013}
{Contreras}, S., {Baugh}, C.~M., {Norberg}, P., \& {Padilla}, N. 2013, \mnras, 432, 2717

\bibitem[{{Contreras} {et~al.}(2015){Contreras}, {Baugh}, {Norberg}, \& {Padilla}}]{Contreras2015}
{Contreras}, S., {Baugh}, C.~M., {Norberg}, P., \& {Padilla}, N. 2015, \mnras, 452, 1861

\bibitem[{{Contreras} {et~al.}(2021{\natexlab{c}}){Contreras}, {Chaves-Montero}, {Zennaro}, \& {Angulo}}]{Contreras2021_cosmo}
{Contreras}, S., {Chaves-Montero}, J., {Zennaro}, M., \& {Angulo}, R.~E. 2021{\natexlab{c}}, \mnras, 507, 3412

\bibitem[{{Contreras} {et~al.}(2017){Contreras}, {Zehavi}, {Baugh}, {Padilla}, \& {Norberg}}]{Contreras2017}
{Contreras}, S., {Zehavi}, I., {Baugh}, C.~M., {Padilla}, N., \& {Norberg}, P. 2017, \mnras, 465, 2833

\bibitem[{{Contreras} {et~al.}(2019){Contreras}, {Zehavi}, {Padilla}, {Baugh}, {Jim{\'e}nez}, \& {Lacerna}}]{Contreras2019}
{Contreras}, S., {Zehavi}, I., {Padilla}, N., {et~al.} 2019, \mnras, 484, 1133

\bibitem[{{Dalal} {et~al.}(2008){Dalal}, {White}, {Bond}, \& {Shirokov}}]{Dalal2008}
{Dalal}, N., {White}, M., {Bond}, J.~R., \& {Shirokov}, A. 2008, \apj, 687, 12

\bibitem[{{Davis} {et~al.}(1985){Davis}, {Efstathiou}, {Frenk}, \& {White}}]{Davis1985}
{Davis}, M., {Efstathiou}, G., {Frenk}, C.~S., \& {White}, S.~D.~M. 1985, \apj, 292, 371

\bibitem[{{Dolag} {et~al.}(2009){Dolag}, {Borgani}, {Murante}, \& {Springel}}]{Dolag2009}
{Dolag}, K., {Borgani}, S., {Murante}, G., \& {Springel}, V. 2009, \mnras, 399, 497

\bibitem[{{Dragomir} {et~al.}(2018){Dragomir}, {Rodr{\'\i}guez-Puebla}, {Primack}, \& {Lee}}]{Dragomir2018}
{Dragomir}, R., {Rodr{\'\i}guez-Puebla}, A., {Primack}, J.~R., \& {Lee}, C.~T. 2018, \mnras, 476, 741

\bibitem[{{Faltenbacher} \& {White}(2010)}]{faltenbacher2010}
{Faltenbacher}, A. \& {White}, S. D.~M. 2010, \apj, 708, 469

\bibitem[{{Favole} {et~al.}(2022){Favole}, {Montero-Dorta}, {Artale}, {Contreras}, {Zehavi}, \& {Xu}}]{Favole2022}
{Favole}, G., {Montero-Dorta}, A.~D., {Artale}, M.~C., {et~al.} 2022, \mnras, 509, 1614

\bibitem[{{Gao} {et~al.}(2005){Gao}, {Springel}, \& {White}}]{gao2005}
{Gao}, L., {Springel}, V., \& {White}, S.~D.~M. 2005, \mnras, 363, L66

\bibitem[{{Gao} \& {White}(2007)}]{Gao2007}
{Gao}, L. \& {White}, S.~D.~M. 2007, \mnras, 377, L5

\bibitem[{{Guo} {et~al.}(2016){Guo}, {Zheng}, {Behroozi}, {Zehavi}, {Chuang}, {Comparat}, {Favole}, {Gottloeber}, {Klypin}, {Prada}, {Rodr{\'\i}guez-Torres}, {Weinberg}, \& {Yepes}}]{Guo2016}
{Guo}, H., {Zheng}, Z., {Behroozi}, P.~S., {et~al.} 2016, \mnras, 459, 3040

\bibitem[{{Guo} {et~al.}(2015){Guo}, {Zheng}, {Zehavi}, {Behroozi}, {Chuang}, {Comparat}, {Favole}, {Gottloeber}, {Klypin}, {Prada}, {Weinberg}, \& {Yepes}}]{Guo2015}
{Guo}, H., {Zheng}, Z., {Zehavi}, I., {et~al.} 2015, \mnras, 453, 4368

\bibitem[{{Hadzhiyska} {et~al.}(2021){Hadzhiyska}, {Bose}, {Eisenstein}, \& {Hernquist}}]{Hadzhiyska2021}
{Hadzhiyska}, B., {Bose}, S., {Eisenstein}, D., \& {Hernquist}, L. 2021, \mnras, 501, 1603

\bibitem[{{Hahn} {et~al.}(2009){Hahn}, {Porciani}, {Dekel}, \& {Carollo}}]{Hahn2009}
{Hahn}, O., {Porciani}, C., {Dekel}, A., \& {Carollo}, C.~M. 2009, \mnras, 398, 1742

\bibitem[{{Han} {et~al.}(2019){Han}, {Li}, {Jing}, {Nishimichi}, {Wang}, \& {Jiang}}]{Han2018}
{Han}, J., {Li}, Y., {Jing}, Y., {et~al.} 2019, \mnras, 482, 1900

\bibitem[{{Hearin} {et~al.}(2015){Hearin}, {Watson}, \& {van den Bosch}}]{Hearin2015}
{Hearin}, A.~P., {Watson}, D.~F., \& {van den Bosch}, F.~C. 2015, \mnras, 452, 1958

\bibitem[{{Jing} {et~al.}(1998){Jing}, {Mo}, \& {B{\"o}rner}}]{Jing1998}
{Jing}, Y.~P., {Mo}, H.~J., \& {B{\"o}rner}, G. 1998, \apj, 494, 1

\bibitem[{{Johnson} {et~al.}(2019){Johnson}, {Maller}, {Berlind}, {Sinha}, \& {Holley-Bockelmann}}]{Johnson2019}
{Johnson}, J.~W., {Maller}, A.~H., {Berlind}, A.~A., {Sinha}, M., \& {Holley-Bockelmann}, J.~K. 2019, \mnras, 486, 1156

\bibitem[{{Kaiser}(1984)}]{Kaiser1984}
{Kaiser}, N. 1984, \apjl, 284, L9

\bibitem[{{Lazeyras} {et~al.}(2017){Lazeyras}, {Musso}, \& {Schmidt}}]{Lazeyras2017}
{Lazeyras}, T., {Musso}, M., \& {Schmidt}, F. 2017, \jcap, 2017, 059

\bibitem[{{Lehmann} {et~al.}(2017){Lehmann}, {Mao}, {Becker}, {Skillman}, \& {Wechsler}}]{Lehmann2017}
{Lehmann}, B.~V., {Mao}, Y.-Y., {Becker}, M.~R., {Skillman}, S.~W., \& {Wechsler}, R.~H. 2017, \apj, 834, 37

\bibitem[{{Li} {et~al.}(2008){Li}, {Mo}, \& {Gao}}]{Li2008}
{Li}, Y., {Mo}, H.~J., \& {Gao}, L. 2008, \mnras, 389, 1419

\bibitem[{{Mao} {et~al.}(2018){Mao}, {Zentner}, \& {Wechsler}}]{Mao2018}
{Mao}, Y.-Y., {Zentner}, A.~R., \& {Wechsler}, R.~H. 2018, \mnras, 474, 5143

\bibitem[{{Marinacci} {et~al.}(2018){Marinacci}, {Vogelsberger}, {Pakmor}, {Torrey}, {Springel}, {Hernquist}, {Nelson}, {Weinberger}, {Pillepich}, {Naiman}, \& {Genel}}]{Marinacci2018}
{Marinacci}, F., {Vogelsberger}, M., {Pakmor}, R., {et~al.} 2018, \mnras, 480, 5113

\bibitem[{{Mo} \& {White}(1996)}]{Mo1996}
{Mo}, H.~J. \& {White}, S.~D.~M. 1996, \mnras, 282, 347

\bibitem[{{Montero-Dorta} {et~al.}(2020){Montero-Dorta}, {Artale}, {Abramo}, {Tucci}, {Padilla}, {Sato-Polito}, {Lacerna}, {Rodriguez}, \& {Angulo}}]{MonteroDorta2020B}
{Montero-Dorta}, A.~D., {Artale}, M.~C., {Abramo}, L.~R., {et~al.} 2020, \mnras, 496, 1182

\bibitem[{{Montero-Dorta} {et~al.}(2021){Montero-Dorta}, {Chaves-Montero}, {Artale}, \& {Favole}}]{MonteroDorta2021}
{Montero-Dorta}, A.~D., {Chaves-Montero}, J., {Artale}, M.~C., \& {Favole}, G. 2021, \mnras, 508, 940

\bibitem[{{Montero-Dorta} \& {Rodriguez}(2024)}]{MonteroRodriguez2024}
{Montero-Dorta}, A.~D. \& {Rodriguez}, F. 2024, \mnras, 531, 290

\bibitem[{{Musso} {et~al.}(2018){Musso}, {Cadiou}, {Pichon}, {Codis}, {Kraljic}, \& {Dubois}}]{Musso2018}
{Musso}, M., {Cadiou}, C., {Pichon}, C., {et~al.} 2018, \mnras, 476, 4877

\bibitem[{{Naiman} {et~al.}(2018){Naiman}, {Pillepich}, {Springel}, {Ramirez-Ruiz}, {Torrey}, {Vogelsberger}, {Pakmor}, {Nelson}, {Marinacci}, {Hernquist}, {Weinberger}, \& {Genel}}]{Naiman2018}
{Naiman}, J.~P., {Pillepich}, A., {Springel}, V., {et~al.} 2018, \mnras, 477, 1206

\bibitem[{{Navarro} {et~al.}(1997){Navarro}, {Frenk}, \& {White}}]{nfw1997}
{Navarro}, J.~F., {Frenk}, C.~S., \& {White}, S. D.~M. 1997, \apj, 490, 493

\bibitem[{{Nelson} {et~al.}(2018){Nelson}, {Pillepich}, {Springel}, {Weinberger}, {Hernquist}, {Pakmor}, {Genel}, {Torrey}, {Vogelsberger}, {Kauffmann}, {Marinacci}, \& {Naiman}}]{Nelson2018_ColorBim}
{Nelson}, D., {Pillepich}, A., {Springel}, V., {et~al.} 2018, \mnras, 475, 624

\bibitem[{{Nelson} {et~al.}(2019){Nelson}, {Springel}, {Pillepich}, {Rodriguez-Gomez}, {Torrey}, {Genel}, {Vogelsberger}, {Pakmor}, {Marinacci}, {Weinberger}, {Kelley}, {Lovell}, {Diemer}, \& {Hernquist}}]{Nelson2019}
{Nelson}, D., {Springel}, V., {Pillepich}, A., {et~al.} 2019, Computational Astrophysics and Cosmology, 6, 2

\bibitem[{{Paranjape} {et~al.}(2018){Paranjape}, {Hahn}, \& {Sheth}}]{Paranjape2018}
{Paranjape}, A., {Hahn}, O., \& {Sheth}, R.~K. 2018, \mnras, 476, 3631

\bibitem[{{Peacock} \& {Smith}(2000)}]{Peacock2000}
{Peacock}, J.~A. \& {Smith}, R.~E. 2000, \mnras, 318, 1144

\bibitem[{{Pillepich} {et~al.}(2018{\natexlab{a}}){Pillepich}, {Nelson}, {Hernquist}, {Springel}, {Pakmor}, {Torrey}, {Weinberger}, {Genel}, {Naiman}, {Marinacci}, \& {Vogelsberger}}]{Pillepich2018b}
{Pillepich}, A., {Nelson}, D., {Hernquist}, L., {et~al.} 2018{\natexlab{a}}, \mnras, 475, 648

\bibitem[{{Pillepich} {et~al.}(2018{\natexlab{b}}){Pillepich}, {Springel}, {Nelson}, {Genel}, {Naiman}, {Pakmor}, {Hernquist}, {Torrey}, {Vogelsberger}, {Weinberger}, \& {Marinacci}}]{Pillepich2018}
{Pillepich}, A., {Springel}, V., {Nelson}, D., {et~al.} 2018{\natexlab{b}}, \mnras, 473, 4077

\bibitem[{{Planck Collaboration} {et~al.}(2016){Planck Collaboration}, {Ade}, {Aghanim}, {Arnaud}, {Ashdown}, {Aumont}, {Baccigalupi}, {Banday}, {Barreiro}, {Bartlett}, {Bartolo}, {Battaner}, {Battye}, {Benabed}, {Beno{\^\i}t}, {Benoit-L{\'e}vy}, {Bernard}, {Bersanelli}, {Bielewicz}, {Bock}, {Bonaldi}, {Bonavera}, {Bond}, {Borrill}, {Bouchet}, {Boulanger}, {Bucher}, {Burigana}, {Butler}, {Calabrese}, {Cardoso}, {Catalano}, {Challinor}, {Chamballu}, {Chary}, {Chiang}, {Chluba}, {Christensen}, {Church}, {Clements}, {Colombi}, {Colombo}, {Combet}, {Coulais}, {Crill}, {Curto}, {Cuttaia}, {Danese}, {Davies}, {Davis}, {de Bernardis}, {de Rosa}, {de Zotti}, {Delabrouille}, {D{\'e}sert}, {Di Valentino}, {Dickinson}, {Diego}, {Dolag}, {Dole}, {Donzelli}, {Dor{\'e}}, {Douspis}, {Ducout}, {Dunkley}, {Dupac}, {Efstathiou}, {Elsner}, {En{\ss}lin}, {Eriksen}, {Farhang}, {Fergusson}, {Finelli}, {Forni}, {Frailis}, {Fraisse}, {Franceschi}, {Frejsel}, {Galeotta}, {Galli}, {Ganga}, {Gauthier}, {Gerbino}, {Ghosh}, {Giard},
  {Giraud-H{\'e}raud}, {Giusarma}, {Gjerl{\o}w}, {Gonz{\'a}lez-Nuevo}, {G{\'o}rski}, {Gratton}, {Gregorio}, {Gruppuso}, {Gudmundsson}, {Hamann}, {Hansen}, {Hanson}, {Harrison}, {Helou}, {Henrot-Versill{\'e}}, {Hern{\'a}ndez-Monteagudo}, {Herranz}, {Hildebrand t}, {Hivon}, {Hobson}, {Holmes}, {Hornstrup}, {Hovest}, {Huang}, {Huffenberger}, {Hurier}, {Jaffe}, {Jaffe}, {Jones}, {Juvela}, {Keih{\"a}nen}, {Keskitalo}, {Kisner}, {Kneissl}, {Knoche}, {Knox}, {Kunz}, {Kurki-Suonio}, {Lagache}, {L{\"a}hteenm{\"a}ki}, {Lamarre}, {Lasenby}, {Lattanzi}, {Lawrence}, {Leahy}, {Leonardi}, {Lesgourgues}, {Levrier}, {Lewis}, {Liguori}, {Lilje}, {Linden-V{\o}rnle}, {L{\'o}pez-Caniego}, {Lubin}, {Mac{\'\i}as-P{\'e}rez}, {Maggio}, {Maino}, {Mandolesi}, {Mangilli}, {Marchini}, {Maris}, {Martin}, {Martinelli}, {Mart{\'\i}nez-Gonz{\'a}lez}, {Masi}, {Matarrese}, {McGehee}, {Meinhold}, {Melchiorri}, {Melin}, {Mendes}, {Mennella}, {Migliaccio}, {Millea}, {Mitra}, {Miville-Desch{\^e}nes}, {Moneti}, {Montier}, {Morgante}, {Mortlock},
  {Moss}, {Munshi}, {Murphy}, {Naselsky}, {Nati}, {Natoli}, {Netterfield}, {N{\o}rgaard-Nielsen}, {Noviello}, {Novikov}, {Novikov}, {Oxborrow}, {Paci}, {Pagano}, {Pajot}, {Paladini}, {Paoletti}, {Partridge}, {Pasian}, {Patanchon}, {Pearson}, {Perdereau}, {Perotto}, {Perrotta}, {Pettorino}, {Piacentini}, {Piat}, {Pierpaoli}, {Pietrobon}, {Plaszczynski}, {Pointecouteau}, {Polenta}, {Popa}, {Pratt}, {Pr{\'e}zeau}, {Prunet}, {Puget}, {Rachen}, {Reach}, {Rebolo}, {Reinecke}, {Remazeilles}, {Renault}, {Renzi}, {Ristorcelli}, {Rocha}, {Rosset}, {Rossetti}, {Roudier}, {Rouill{\'e} d'Orfeuil}, {Rowan-Robinson}, {Rubi{\~n}o-Mart{\'\i}n}, {Rusholme}, {Said}, {Salvatelli}, {Salvati}, {Sandri}, {Santos}, {Savelainen}, {Savini}, {Scott}, {Seiffert}, {Serra}, {Shellard}, {Spencer}, {Spinelli}, {Stolyarov}, {Stompor}, {Sudiwala}, {Sunyaev}, {Sutton}, {Suur-Uski}, {Sygnet}, {Tauber}, {Terenzi}, {Toffolatti}, {Tomasi}, {Tristram}, {Trombetti}, {Tucci}, {Tuovinen}, {T{\"u}rler}, {Umana}, {Valenziano}, {Valiviita}, {Van Tent},
  {Vielva}, {Villa}, {Wade}, {Wandelt}, {Wehus}, {White}, {White}, {Wilkinson}, {Yvon}, {Zacchei}, \& {Zonca}}]{Planck2016}
{Planck Collaboration}, {Ade}, P.~A.~R., {Aghanim}, N., {et~al.} 2016, \aap, 594, A13

\bibitem[{{Ramakrishnan} {et~al.}(2019){Ramakrishnan}, {Paranjape}, {Hahn}, \& {Sheth}}]{Ramakrishnan2019}
{Ramakrishnan}, S., {Paranjape}, A., {Hahn}, O., \& {Sheth}, R.~K. 2019, \mnras, 489, 2977

\bibitem[{{Reddick} {et~al.}(2013){Reddick}, {Wechsler}, {Tinker}, \& {Behroozi}}]{Reddick2013}
{Reddick}, R.~M., {Wechsler}, R.~H., {Tinker}, J.~L., \& {Behroozi}, P.~S. 2013, \apj, 771, 30

\bibitem[{{Salcedo} {et~al.}(2018){Salcedo}, {Maller}, {Berlind}, {Sinha}, {McBride}, {Behroozi}, {Wechsler}, \& {Weinberg}}]{2018Salcedo}
{Salcedo}, A.~N., {Maller}, A.~H., {Berlind}, A.~A., {et~al.} 2018, \mnras, 475, 4411

\bibitem[{{Sato-Polito} {et~al.}(2019){Sato-Polito}, {Montero-Dorta}, {Abramo}, {Prada}, \& {Klypin}}]{SatoPolito2019}
{Sato-Polito}, G., {Montero-Dorta}, A.~D., {Abramo}, L.~R., {Prada}, F., \& {Klypin}, A. 2019, \mnras, 487, 1570

\bibitem[{{Sheth} {et~al.}(2001){Sheth}, {Mo}, \& {Tormen}}]{Sheth2001}
{Sheth}, R.~K., {Mo}, H.~J., \& {Tormen}, G. 2001, \mnras, 323, 1

\bibitem[{{Sheth} \& {Tormen}(1999)}]{ShethTormen1999}
{Sheth}, R.~K. \& {Tormen}, G. 1999, \mnras, 308, 119

\bibitem[{{Sheth} \& {Tormen}(2004)}]{Sheth2004}
{Sheth}, R.~K. \& {Tormen}, G. 2004, \mnras, 350, 1385

\bibitem[{{Springel} {et~al.}(2018){Springel}, {Pakmor}, {Pillepich}, {Weinberger}, {Nelson}, {Hernquist}, {Vogelsberger}, {Genel}, {Torrey}, {Marinacci}, \& {Naiman}}]{Springel2018}
{Springel}, V., {Pakmor}, R., {Pillepich}, A., {et~al.} 2018, \mnras, 475, 676

\bibitem[{{Springel} {et~al.}(2001){Springel}, {White}, {Tormen}, \& {Kauffmann}}]{Springel2001}
{Springel}, V., {White}, S. D.~M., {Tormen}, G., \& {Kauffmann}, G. 2001, \mnras, 328, 726

\bibitem[{{Tinker} {et~al.}(2010){Tinker}, {Robertson}, {Kravtsov}, {Klypin}, {Warren}, {Yepes}, \& {Gottl{\"o}ber}}]{Tinker2010}
{Tinker}, J.~L., {Robertson}, B.~E., {Kravtsov}, A.~V., {et~al.} 2010, \apj, 724, 878

\bibitem[{{Tucci} {et~al.}(2021){Tucci}, {Montero-Dorta}, {Abramo}, {Sato-Polito}, \& {Artale}}]{Tucci2021}
{Tucci}, B., {Montero-Dorta}, A.~D., {Abramo}, L.~R., {Sato-Polito}, G., \& {Artale}, M.~C. 2021, \mnras, 500, 2777

\bibitem[{{Vale} \& {Ostriker}(2006)}]{Vale2006}
{Vale}, A. \& {Ostriker}, J.~P. 2006, \mnras, 371, 1173

\bibitem[{{Wechsler} {et~al.}(2006){Wechsler}, {Zentner}, {Bullock}, {Kravtsov}, \& {Allgood}}]{Wechsler2006}
{Wechsler}, R.~H., {Zentner}, A.~R., {Bullock}, J.~S., {Kravtsov}, A.~V., \& {Allgood}, B. 2006, \apj, 652, 71

\bibitem[{{Xu} {et~al.}(2021{\natexlab{a}}){Xu}, {Kumar}, {Zehavi}, \& {Contreras}}]{Xu2021a}
{Xu}, X., {Kumar}, S., {Zehavi}, I., \& {Contreras}, S. 2021{\natexlab{a}}, \mnras, 507, 4879

\bibitem[{{Xu} {et~al.}(2021{\natexlab{b}}){Xu}, {Zehavi}, \& {Contreras}}]{Xu2021b}
{Xu}, X., {Zehavi}, I., \& {Contreras}, S. 2021{\natexlab{b}}, \mnras, 502, 3242

\bibitem[{{Zheng} {et~al.}(2005){Zheng}, {Berlind}, {Weinberg}, {Benson}, {Baugh}, {Cole}, {Dav{\'e}}, {Frenk}, {Katz}, \& {Lacey}}]{Zheng2005}
{Zheng}, Z., {Berlind}, A.~A., {Weinberg}, D.~H., {et~al.} 2005, \apj, 633, 791

\bibitem[{{Zheng} {et~al.}(2007){Zheng}, {Coil}, \& {Zehavi}}]{Zheng2007}
{Zheng}, Z., {Coil}, A.~L., \& {Zehavi}, I. 2007, \apj, 667, 760

\end{thebibliography}

\end{document}